\documentclass[a4paper,fleqn]{cas-dc}

\usepackage[numbers,sort&compress]{natbib}

\usepackage{natbib}

\usepackage[capitalise]{cleveref}
\creflabelformat{equation}{#2\textup{#1}#3}
\Crefname{equation}{Eq.}{Eqs.}
\Crefname{figure}{Fig.}{Figs.}

\newcommand{\urad}{\mu\mathrm{m}\cdot\mathrm{rad}}
\newcommand{\nrad}{\mathrm{nm}\cdot\mathrm{rad}}
\begin{document}
\let\WriteBookmarks\relax
\def\floatpagepagefraction{1}
\def\textpagefraction{.001}

\shorttitle{}    

\shortauthors{}  

\title [mode = title]{Limits on energy recovery for positron acceleration in a warm plasma column}  




\author[1]{Max Varverakis}[orcid=0000-0003-4338-924X,]
\cormark[1]
\credit{Software, Formal analysis, Investigation, Data Curation, Visualization, Writing - Original Draft, Writing - Review \& Editing}

% \cormark[1]

\fnmark[1]

\ead{maxvarverakis@physics.ucla.edu}



\affiliation[1]{organization={{SLAC} National Accelerator Laboratory},
            addressline={2575 Sand Hill Road}, 
            city={Menlo Park},
            % citysep={}, % Uncomment if no comma needed between city and postcode
            postcode={94025}, 
            state={CA},
            country={USA}}

\author[1]{Spencer Gessner}[orcid=0000-0002-9713-1116]
\cormark[2]
\credit{Conceptualization, Supervision, Project administration, Writing - Review \& Editing, Funding acquisition}

% Footnote of the second author

% Email id of the second author
\ead{sgess@slac.stanford.edu}


% \affiliation[2]{organization={},
% %          citysep={}, % Uncomment if no comma needed between city and postcode

\author[2]{Severin Diederichs}[orcid=0000-0001-9079-0461]
\affiliation[2]{organization={Deutsches Elektronen-Synchroton {DESY}},
            addressline={Notkestraße 85}, 
            city={Hamburg},
            % citysep={}, % Uncomment if no comma needed between city and postcode
            postcode={22607}, 
            country={Germany}}
\credit{Conceptualization, Software, Resources, Writing - Review \& Editing}
\fnmark[2]

\author[3]{Carl A. Lindstr{\o}m}[orcid=0000-0002-0769-0425]
\affiliation[3]{organization={Department of Physics, University of Oslo},
            addressline={PO Box 1048 Blindern}, 
            city={Oslo},
            % citysep={}, % Uncomment if no comma needed between city and postcode
            postcode={N-0316}, 
            country={Norway}}
\credit{Resources, Writing - Review \& Editing, Funding acquisition}

\cortext[1]{Corresponding author}
\cortext[2]{Principal corresponding author}

\fntext[1]{Present address: Department of Physics \& Astronomy, University of California, Los Angeles, 90095, CA, USA}
\fntext[2]{Present address: CERN, Espl. des Particules 1, 1211 Geneva, Switzerland}


\begin{abstract}
Beam-driven plasma wakefield acceleration (PWFA) is a promising candidate for realizing a future linear collider, owing to the extremely high field gradients that are produced in plasmas. Despite recent success in high-gradient and high-efficiency electron PWFA, efficient positron PWFA remains a challenge. The warm plasma column regime offers stable, high-quality acceleration of positron beams in a plasma electron filament, albeit at low energy-transfer efficiencies. We explore the possibility of employing an electron energy recovery beam to extract residual energy from the plasma wakefield and increase the efficiency of this scheme. Our particle-in-cell simulations reveal that certain mechanisms, including higher-order ionization and ion motion, limit our ability to further extract energy from the plasma wakefield.
% there is a trade-off between beam quality and efficiency which limits our ability to further extract energy from the plasma wakefield.
%which implies a limit on the efficiency of a future electron-positron linear collider that utilizes the plasma column regime for positron acceleration.
\end{abstract}


% \begin{highlights}
% \item 
% \item 
% \item 
% \end{highlights}

%\nocite{*}

% Keywords

\maketitle

% Main text
\section{Introduction}\label{sec:intro}
The US High Energy Physics community aims to study physics at the 10 TeV parton center-of-momentum (pCM) scale~\cite{Gessner:2025,P5:2023,Roser2023}.
A possible path to this energy frontier lies in plasma-based acceleration technology because of the high accelerating gradients that can be produced~\cite{Rosenzweig1991, Chen2020, Schroeder2010, Delahaye2014, Schroeder2023}. In particular, beam-driven plasma wakefield acceleration (PWFA) addresses many challenges of realizing a TeV-scale linear collider~\cite{Blumenfeld2007, Litos2014}.

Optimal beam loading in the nonlinear plasma (blowout) regime can result in high-efficiency and high-quality PWFA of electrons~\cite{Tzoufras2008} and has been demonstrated experimentally~\cite{Lindstrom2021,Lindstrom2024}. However, the extension of stable and efficient PWFA to positrons has proven to be difficult and is an active area of research~\cite{Hogan2003, Blue2003, Muggli2008, Corde2015, Doche2017, Lee2001}. It has been found that the blowout regime can create high-density regions of plasma electrons, providing the necessary focusing and acceleration fields for positron beams.

Several plasma configurations have been proposed to achieve positron acceleration in the blowout regime utilizing dense regions of plasma electrons, known as electron filaments. Such schemes include plasma columns~\cite{Diederichs2019, Diederichs2020, Diederichs2022, Diederichs2022_2}, blowout wake tail loading~\cite{Lotov2007,Zhou2022,Liu2023,Wang2021}, hollow plasma channels~\cite{Silva2021,Zhou2021,Gessner2023}, and positively charged annular drive beams~\cite{Han2026}. However, the nonlinear transverse focusing fields and longitudinally varying accelerating fields produced in a cold plasma electron filament can lead to beam degradation in the form of energy spread and/or emittance growth.

Recent work in the plasma column regime has shown that warm plasma generates an elongated electron filament with partially linear focusing fields and limited longitudinal variation of the accelerating fields~\cite{Diederichs2023}. For sufficiently small positron beams, warm plasma electron filaments can provide self-stable, high-gradient positron acceleration while preserving emittance and minimizing energy spread~\cite{Diederichs2025}.

Despite the promise of the warm plasma column, the energy-transfer efficiency from the drive beam to the positron beam has not yet surpassed 5\% due to beam loading limitations~\cite{Diederichs2020,Varverakis2025}. In addition to reaching high energy collisions, a future linear collider must achieve high luminosities, wherein luminosity-per-beam power, a key figure of merit, is proportional to energy-transfer efficiency $\eta$:
\begin{equation}
    \frac{\mathcal{L}}{P_{\mathrm{tot}}} = \frac{\eta N}{4\pi\sigma_x\sigma_y E_b},
\end{equation}
for $N$ the number of particles, $\sigma_{x}$ ($\sigma_{y}$) the RMS beam size in $x$ ($y$), and $E_b$ the beam energy~\cite{Cao2024}.

The energy footprint of particle accelerators is identified as a critical consideration for future collider facilities~\cite{Roser2022,Roser2023}. To minimize the energy consumption of a plasma-based accelerator, it is clear that maximizing the energy-transfer efficiency is paramount. The concept of recovering unused energy from a plasma wake has been investigated in the laser-driven case~\cite{Schroeder2016,Cowley2017}, as well as in the context of energy-recovery linacs (ERLs) utilizing RF deceleration of beams~\cite{ERLReview,Litvinenko2020,EuRoadMap}.

In the warm plasma column regime, it is possible to increase the energy-transfer efficiency beyond the 5\% threshold with the addition of a second electron beam whose purpose is to absorb additional energy from the wakefield generated by the drive beam~\cite{Varverakis2025}. The second electron beam serves as an energy recovery beam analogous to the energy recovery laser suggested for laser wakefield acceleration (LWFA)~\cite{Schroeder2016}.

Our previous study on warm plasma column energy recovery schemes demonstrated the possibility of high-gradient positron acceleration with total efficiency reaching up to 27\%, including the electron recovery beam~\cite{Varverakis2025}. However, these studies did not demonstrate preservation of beam quality. In this work, we use the quasi-static particle-in-cell (PIC) code HiPACE++~\cite{HiPACE++} to simulate positron acceleration in a warm plasma column and study the effects of energy recovery beams on beam quality.

Our simulations suggest that positron beam quality suffers as a result of the nonlinear and longitudinally varying transverse fields induced by the recovery beam.
We attempt to address the ion motion that leads to these fields but are unable to fully mitigate their negative effects.
% We attempt to address the underlying mechanisms that degrade positron beam quality but are unable to fully mitigate the negative effects of the varying focusing fields.
These findings may impose restrictions on the versatility of energy recovery beams in a warm plasma column.

\section{Energy recovery in a warm plasma column}\label{sec:energy_recovery_schemes}

% In the warm plasma column regime, plasma electron trajectories are smeared longitudinally, resulting in an electron filament (a dense region of plasma electrons) that supports the acceleration of positron beams with tens to hundreds of pC charge. Although stability and beam quality preservation has been demonstrated in simulations~\cite{Diederichs2020, Diederichs2022_2, Diederichs2025}, the drive-to-witness beam energy-transfer efficiency falls short of what is desired for future collider applications~\cite{Cao2024}.

% To improve the energy efficiency of the warm plasma column regime, we proposed loading a trailing electron beam in addition to the positron beam~\cite{Varverakis2025}. The trailing electron beam in such configurations will be referred to as the energy recovery beam. Two locations in the wake provide the necessary fields for loading the energy recovery beam: in front of and behind the positron beam.

The energy recovery schemes presented in Ref. \cite{Varverakis2025} were designed to maximize energy-transfer efficiency. However, efficiency is not the only parameter to optimize in a PWFA scheme. The goal of this work is to maximize acceleration efficiency while minimizing energy spread and emittance growth. 
%Thus, the baseline beam configurations presented here provide lower efficiencies than obtained in~\cite{Varverakis2025}.

Previous work identified that the energy recovery scheme with the largest efficiency involves recovery beam placement in front of the positron beam. Therefore, the ``recovery-in-front'' beam configuration serves as our starting point for the simulations presented here. We simulate with a plasma column of radius 2.5 $k_p^{-1}$ and at a background density of $n_0 = 5 \times 10^{16}$ cm$^{-3}$ and temperature $k_B T = 50$ eV, where $k_p^{-1}$ is the plasma skin depth and $k_B$ is the Boltzmann constant.
In order to highlight the effects of ion motion in the energy recovery regime, our simulations cover two plasma candidates: (1) hydrogen plasma, and (2) argon plasma.
% \textcolor{red}{Mesh refinement statement and/or simulation methods section/appendix?}
%We define the co-moving coordinate in the speed-of-light frame as $\xi = z - ct$. 
We define the energy-transfer efficiency as the ratio of the energy change of the trailing beams to the drive beam:
\begin{equation}
    \eta = -\left( \frac{Q_p\langle E_z\rangle_p + Q_r\langle E_z\rangle_r}{Q_d\langle E_z\rangle_d} \right),\label{eq:efficiency}
\end{equation}
where $Q$ represents the beam charge, $\langle E_z\rangle$ the accelerating field averaged over the beam, and the subscripts $p,r,d$ correspond to the positron, recovery, and drive beams, respectively. 
%In practice, each term in \cref{eq:efficiency} is calculated via the dot product of the per-slice beam charge with the on-axis longitudinal electric field. 
%Off-axis contributions to efficiency are ignored due to the cylindrical symmetry of the blowout regime.

For the simulations presented in this work, we use a $-5.16$ nC bi-Gaussian drive beam with a normalized transverse emittance of $\varepsilon_{x,y}=54.5\ \urad$. All beams are initialized at 5.11 GeV. The witness (positron and recovery) beams are initialized with normalized transverse emittances of $\varepsilon_{x,y}=59.4\ \nrad$ and follow a radial Gaussian for their transverse profiles. These emittances were chosen to be the smallest within the constraints of the simulation resolution. We use the {SALAME} algorithm~\cite{Diederichs2020} to numerically optimize the longitudinal current profiles of the witness beams for the energy recovery simulations, resulting in a flattened average accelerating field across the beams. Note that {SALAME} dictates the final charge of the witness beams.

\begin{figure}[pos=htbp]
    \centering
    \includegraphics[width=0.5\textwidth]{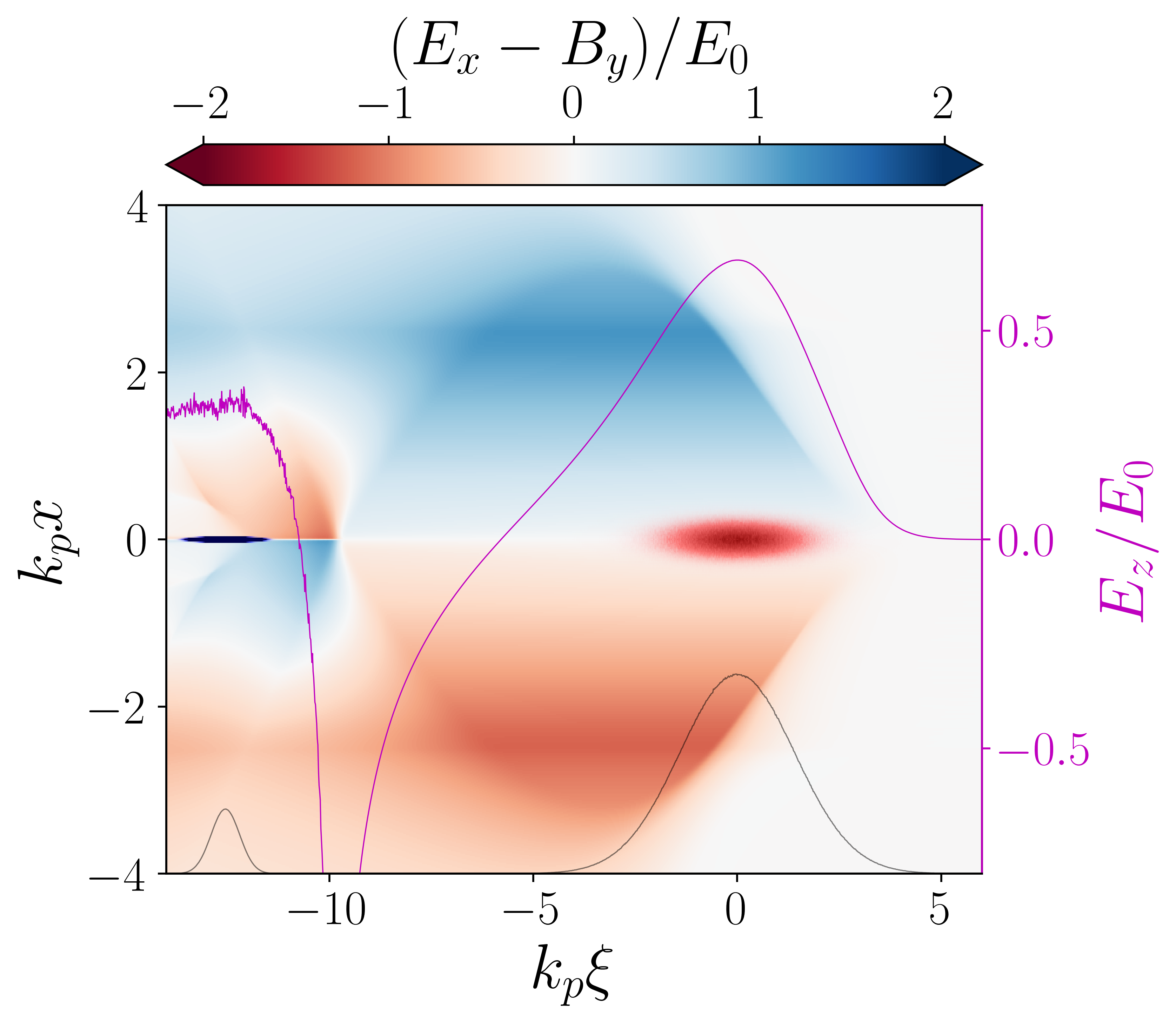}
    \caption{
    % Baseline positron acceleration scheme in a warm plasma column.
    Single-bunch positron acceleration in a warm hydrogen plasma column.
    A $-5.16$ nC electron drive beam accelerates 415 pC of positrons with an energy-transfer efficiency of 4.38~\%. Current densities are displayed in red for electrons and blue for positrons. On-axis longitudinal current profiles are outlined in gray on the horizontal axis, and the on-axis accelerating field is overlaid in magenta. SALAME was not used in this simulation.}
    \label{fig:ExmBy_no_recovery}
\end{figure}

\begin{figure}[pos=h!]
    \centering
    \includegraphics[width=0.5\textwidth]{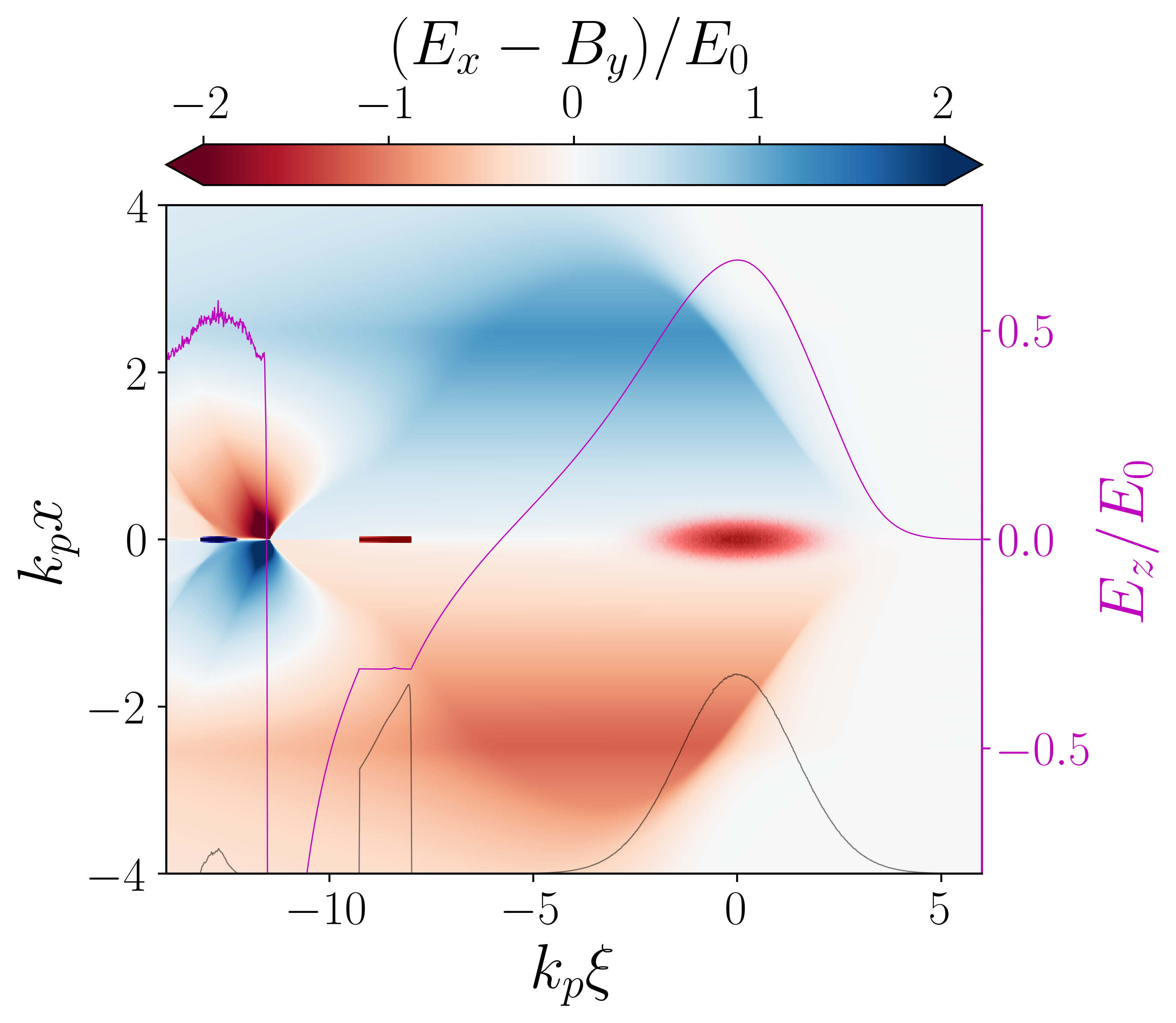}
    \caption{Positron acceleration in a warm hydrogen plasma column with the addition of an electron energy recovery beam. The positron beam has a charge of 91 pC as a result of the longitudinal profile optimization via {SALAME}. The addition of the $-1.39$ nC recovery beam increases the energy-transfer efficiency to 16.8~\%.}
    \label{fig:ExmBy_front_recovery}
\end{figure}

The baseline case for positron acceleration in the warm plasma column regime is presented in \cref{fig:ExmBy_no_recovery}. Adding an energy recovery beam to the baseline case increases the energy-transfer efficiency from 4.38~\% to 16.8~\% and is depicted in \cref{fig:ExmBy_front_recovery}. Despite the desirable outcome of efficiency gains, the addition of the recovery beam exacerbates ion motion in both the recovery beam itself and the positron beam. In the following section, we highlight why ion motion is not desirable for the energy recovery scheme.

\section{Effects of ion motion on beam quality}

\begin{figure}[pos=htbp]
    \centering
    \includegraphics[width=0.5\textwidth]{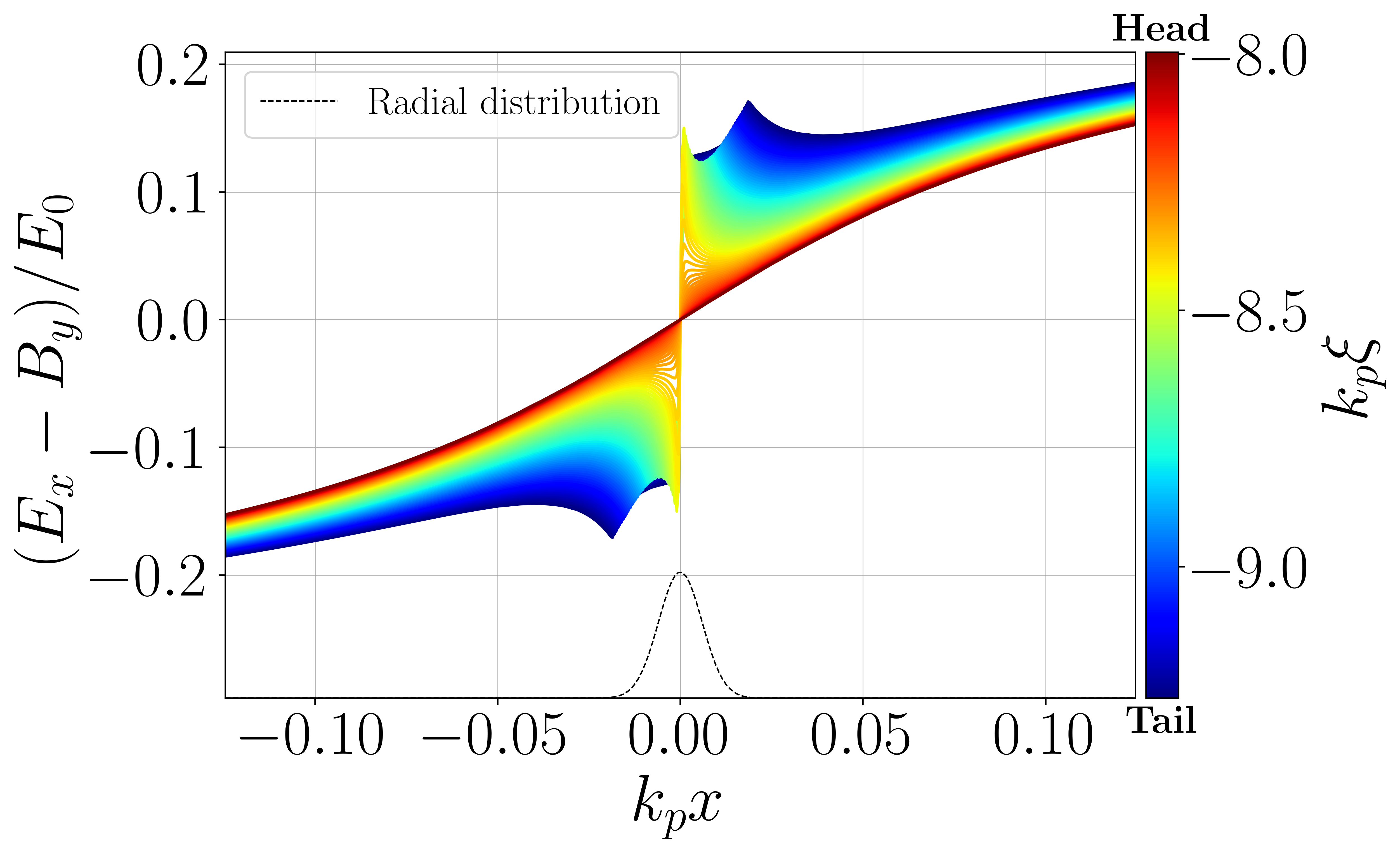}
    \caption{Transverse focusing field across the recovery beam from the simulation presented in \cref{fig:ExmBy_front_recovery}. The dashed black line shows the recovery beam's radial distribution, indicating that the focusing field is nonlinear for the majority of the recovery beam.}
    \label{fig:H_recovery_ion_motion}
\end{figure}

\begin{figure}[pos=htbp]
    \centering
    \includegraphics[width=0.5\textwidth]{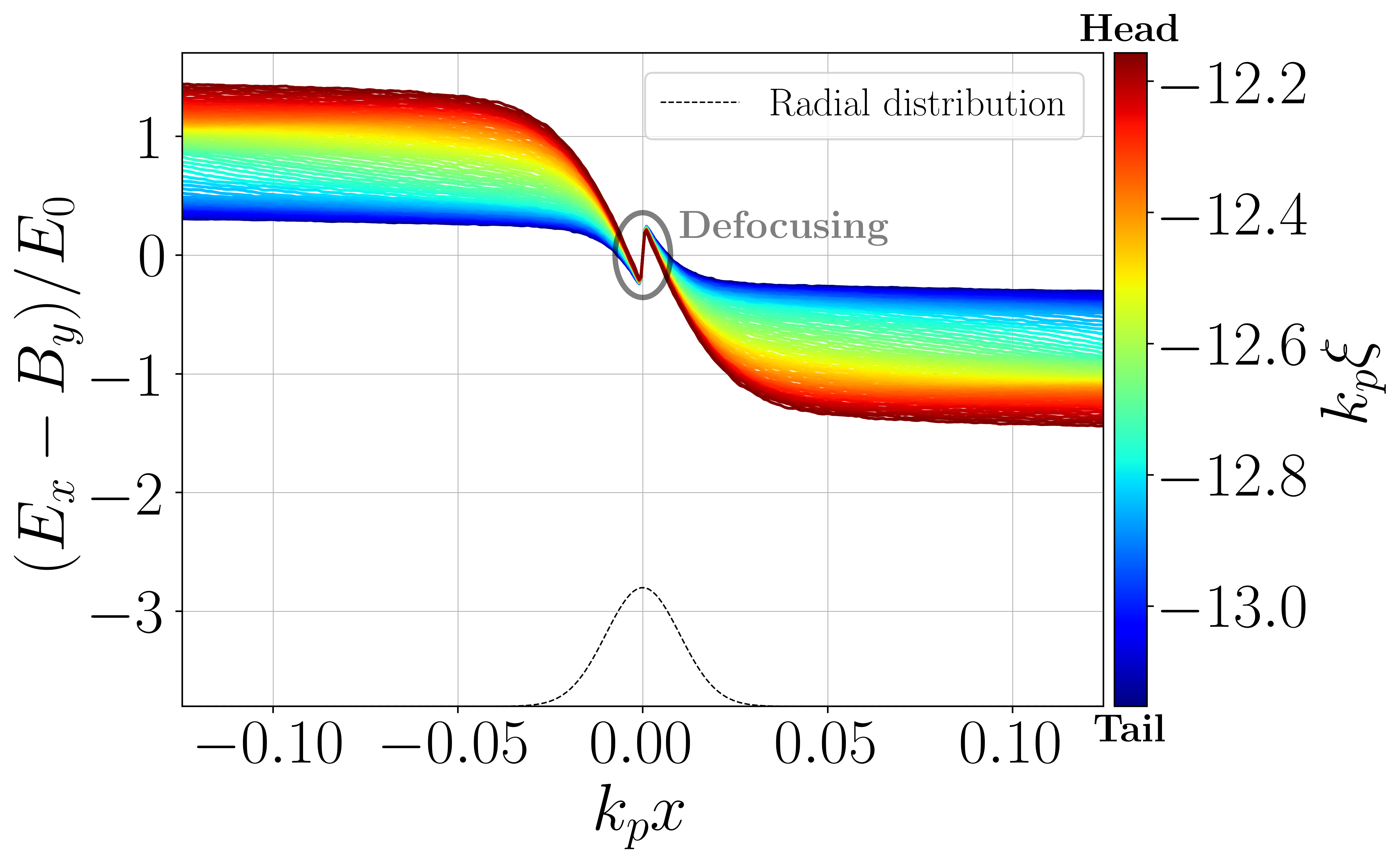}
    \caption{Transverse focusing field across the positron beam from \cref{fig:ExmBy_front_recovery}. The dashed black line shows the positron beam's radial distribution. A defocusing region (circled in gray) is found across the transverse centroid of the positron beam.}
    \label{fig:H_positron_ion_motion}
\end{figure}

Loading dense electron beams in a warm plasma column results in ion motion that creates unfavorable fields for both witness beams. Specifically, the recovery beam experiences nonlinear focusing and the positron beam region becomes defocusing near the beam axis. These field characteristics, illustrated in \cref{fig:H_recovery_ion_motion,fig:H_positron_ion_motion}, lead to significant emittance growth and energy spread.

If we instead use octuply-ionized argon plasma, the ion motion is suppressed by a factor of roughly 8/40 (the charge-to-mass ratio) compared to hydrogen plasma. The ionization level of argon is chosen to prevent higher-order ionization from the recovery beam. This is done to preserve the wakefield structure by maintaining a single plasma ion species.

We can further minimize ion motion by reducing the density of the recovery beam. This can be achieved by increasing the recovery beam's transverse size while holding the total charge roughly constant, at the cost of recovery beam quality. A similar simulation to \cref{fig:ExmBy_front_recovery} was performed with argon plasma and a wide recovery beam, resulting in 13.3~\% efficiency for a $-1.54$ nC recovery beam and an 81 pC positron beam.

\begin{figure}[pos=htbp]
    \centering
    \includegraphics[width=0.5\textwidth]{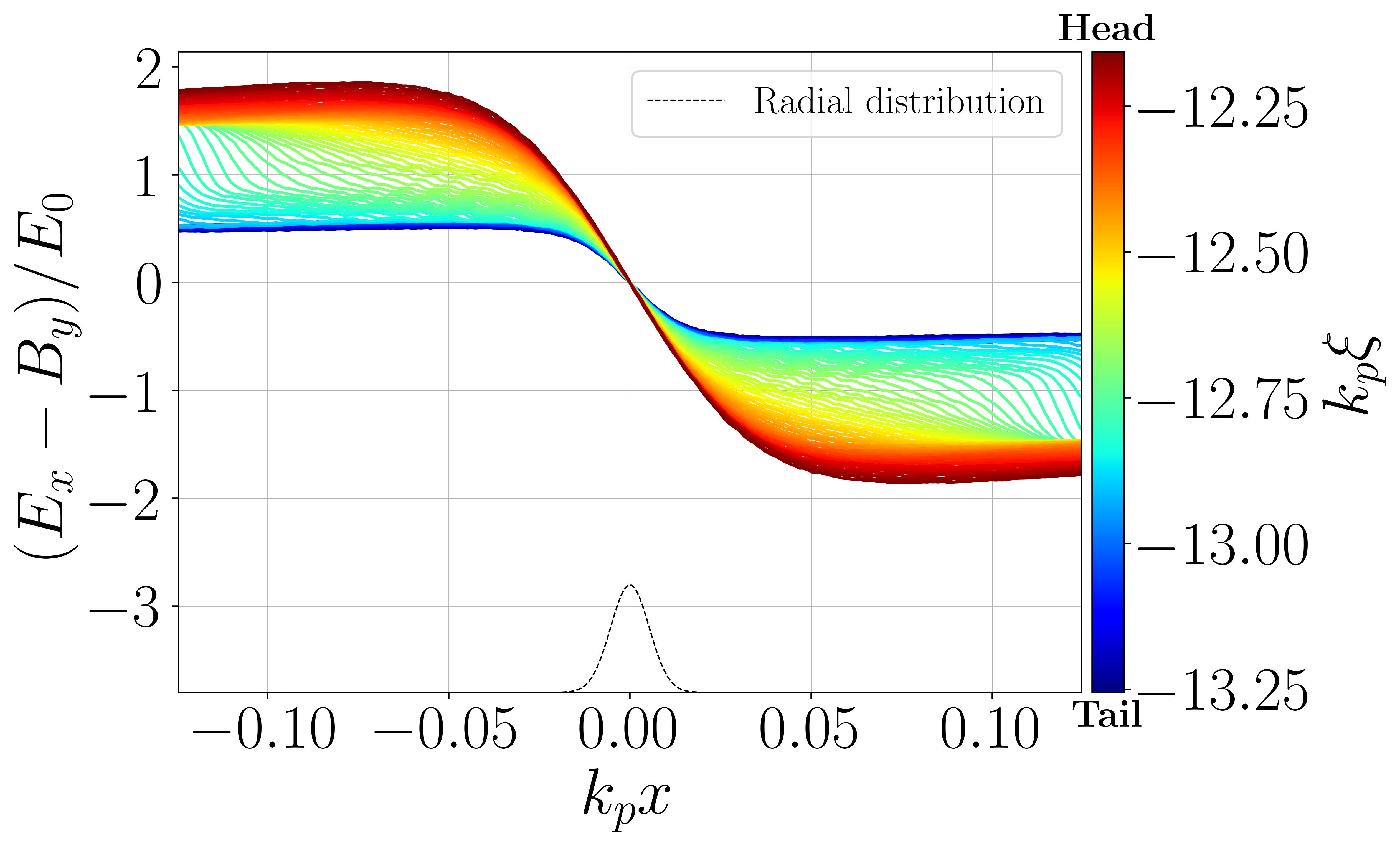}
    \caption{Transverse focusing field across the positron beam from the argon plasma simulation. The defocusing region found in the hydrogen plasma case (\cref{fig:H_positron_ion_motion}) is ameliorated as a result of the suppressed ion motion.}
    \label{fig:Argon_positron_ion_motion}
\end{figure}

The combination of heavy plasma ions and a wide recovery beam sufficiently removes the destabilizing fields for the positron beam that were present in the hydrogen plasma simulation. This is illustrated in \cref{fig:Argon_positron_ion_motion}. Both witness beams see less than a tenth of a percent of energy spread over the course of the beam propagation due to the longitudinally optimized current profiles.

\begin{figure}[pos=htbp]
    \centering
    \includegraphics[width=0.5\textwidth]{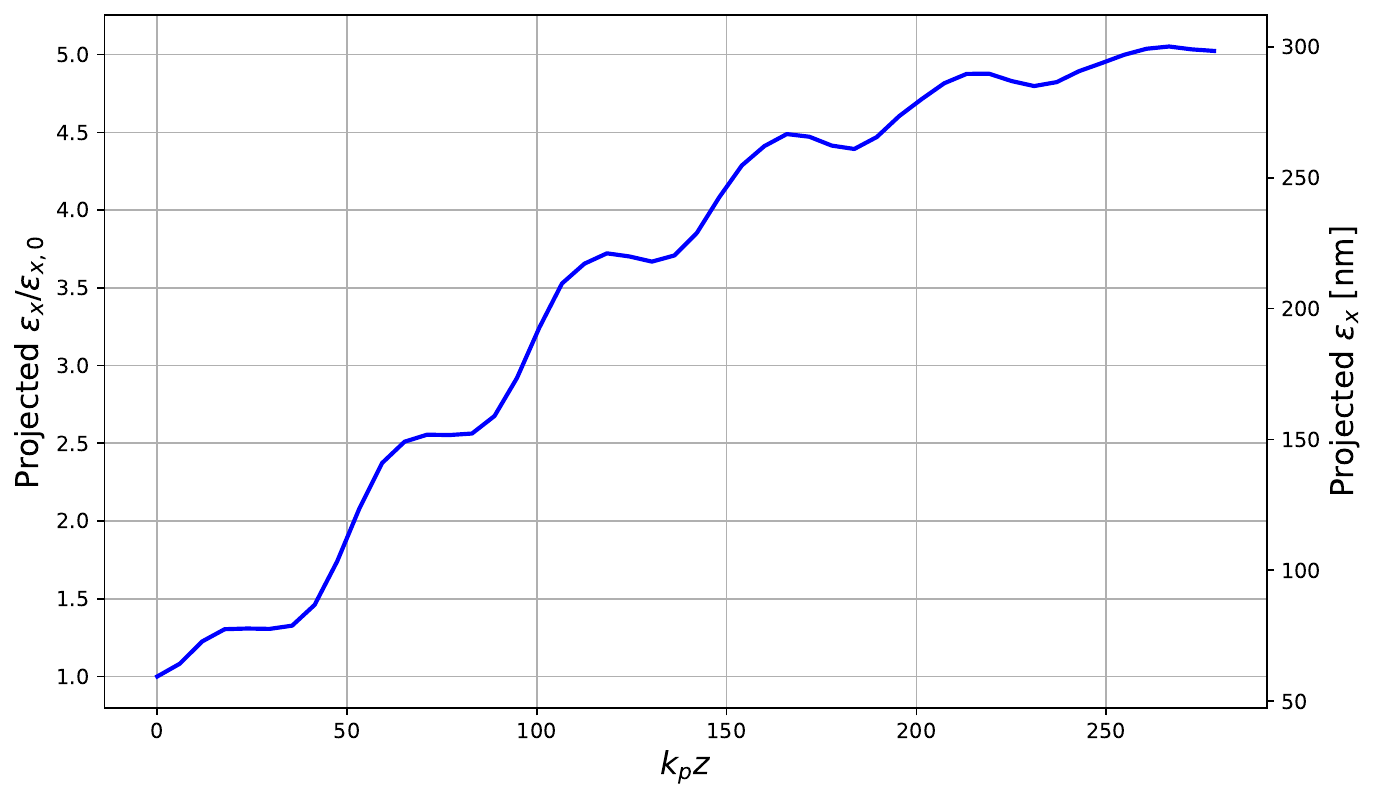}
    \caption{Projected transverse emittance of the positron beam as it propagates roughly 2.96 mm in 9.87 ps with time step $\Delta t = 2.65\ (k_p c)^{-1}$. The left axis indicates the emittance normalized by the initial value.}
    \label{fig:emit_growth}
\end{figure}

Because we simulated beams with longitudinally independent spot sizes, the longitudinal variation in the positron beam's focusing fields led to significant emittance growth, as shown in \cref{fig:emit_growth}. In the absence of slice-by-slice matching, we attempted to numerically quasi-match the positron beam by adjusting the initial beta function of the beam in simulation. However, we were unable to identify a set of parameters that preserve positron beam quality. Further investigation into the energy recovery scheme involving simulations that invoke slice-by-slice matching may reveal whether positron beam emittance can be preserved.

In practice, slice-by-slice matching can be achieved by adiabatically shaping the positron beam profile by loading it at sufficiently low energies~\cite{Benedetti2021}. However, this method may prove difficult since the longitudinal position of the positron beam is constrained to the plasma electron filament. Another method using chromatic focusing may allow per-slice matching of the positron beam, but this has not yet been investigated for the energy recovery schemes covered in this work.

Alternatively, a small enough positron beam can theoretically sample the near-axis transverse fields where longitudinal variation is minimal, allowing the possibility of emittance preservation~\cite{Diederichs2025}. This hypothesis has not been tested because the resolution chosen for our simulations prevents us from initializing smaller beams. Future work with higher resolution simulations could address this as well.

\section{Conclusion}

Efficient, high-gradient acceleration of positron beams in plasma is a prerequisite for future electron-positron colliders based on wakefield technology. We previously demonstrated the ability to extract unused energy from a warm plasma column wakefield to increase the energy-transfer efficiency of the positron acceleration scheme. In this study, we find that there are limitations to energy recovery in a warm plasma column using an electron energy recovery beam due to the nonlinear and longitudinally varying focusing fields that are produced.

Much of the ion motion can be mitigated by selecting a plasma with a sufficiently small charge-to-mass ratio and reducing the recovery beam density. However, the longitudinally varying focusing fields for the positron beam make emittance preservation difficult. A more sophisticated parameter scan of initial beam conditions may identify an operating point that sufficiently matches the positron beam to the wakefield and preserves emittance. Alternatively, higher resolution simulations with smaller positron beams and/or slice-by-slice matching techniques may offer a path towards emittance preservation of positron beams in this energy recovery scheme.

We recognize the possibility of loading the recovery beam behind the positron beam instead of in front. Attempts to load a recovery beam in the second blowout bubble were unsuccessful because the near-axis fields are focusing for positrons throughout its longitudinal extent and hence defocusing for the recovery beam.
Subsequent trailing bubbles have progressively vanishing focusing fields that are not suitable for loading an electron recovery beam.

An alternative energy recovery scheme may consider the use of an energy recovery laser pulse in place of the energy recovery electron bunch. However, the challenges associated with a recovery beam will likely be substituted by comparable challenges related to the nuances of laser pulse recovery.

\section*{Declaration of competing interest}

The authors declare that they have no known competing financial interests or personal relationships that could have appeared to
influence the work reported in this paper.

\section*{Data availability}

All data in this study was obtained via the open-source, 3D, quasi-static PIC code HiPACE++~\cite{HiPACE++},
available on GitHub. The {HiPACE++} input scripts and data analysis files used to produce the results of this work are openly available online~\cite{source_code}, and the simulation data used in this work is publicly available on Zenodo~\cite{varverakis_2026_22017452}.

\section*{Acknowledgments}

The work of M.V. and S.G. is supported by the U.S. Department of Energy under contract DE–AC02–76SF00515. The work of C.A.L. was funded by the European Research Council (ERC Grant Agreement No. 101116161). We acknowledge Sigma2 - the National Infrastructure for High-Performance Computing and Data Storage in Norway for awarding this project access to the LUMI supercomputer, owned by the EuroHPC Joint Undertaking, hosted by CSC (Finland) and the LUMI consortium.
We also gratefully acknowledge the HiPACE++ developers M. Th{\'e}venet and A. Sinn for providing detailed support for using HiPACE++ and simulating on {LUMI}.


% \section{}\label{}

\printcredits

% \bibliographystyle{cas-model2-names}
% \bibliographystyle{ieeetr}
% \bibliographystyle{elsarticle-num}
\bibliographystyle{edited-elsarticle-num}

\bibliography{cas-refs}



\end{document}